\documentclass[english,pra,twocolumn,notitlepage,superscriptaddress,showpacs]{revtex4-1}

\usepackage{graphicx}
\usepackage{color}
\usepackage{amssymb}
\usepackage{amsmath}
\usepackage{setspace}
\usepackage{babel}
\usepackage[utf8x]{inputenc}

\begin{document}

\title{Phonon-induced relaxation and decoherence times of the hybrid qubit in silicon quantum dots}
\author{E. Ferraro}
\email{elena.ferraro@mdm.imm.cnr.it}
\affiliation{CNR-IMM Agrate Unit, Via C. Olivetti 2, 20864 Agrate Brianza (MB), Italy}
\author{M. Fanciulli}
\affiliation{CNR-IMM Agrate Unit, Via C. Olivetti 2, 20864 Agrate Brianza (MB), Italy}
\affiliation{Dipartimento di Scienza dei Materiali, University of Milano Bicocca, Via R. Cozzi, 55, 20126 Milano, Italy}
\author{M. De Michielis}
\email{marco.demichielis@mdm.imm.cnr.it}
\affiliation{CNR-IMM Agrate Unit, Via C. Olivetti 2, 20864 Agrate Brianza (MB), Italy}

\begin{abstract}
We study theoretically the phonon-induced relaxation and decoherence processes in the hybrid qubit in silicon. Hybrid qubit behaves as a charge qubit when the detuning is close to zero and as spin qubit for large detuning values. It is realized starting from an electrostatically defined double quantum dot where three electrons are confined and manipulated through only electrical tuning. By employing a three-level effective model for the qubit and describing the environment bath as a series of harmonic oscillators in the thermal equilibrium states, we extract the relaxation and decoherence times as a function of the bath spectral density and of the bath temperature using the Bloch-Redfield theory. For Si quantum dots the energy dispersion is strongly affected by the physics of the valley, i.e. the conduction band minima, so we also included the contribution of the valley excitations in our analysis. Our results offer fundamental information on the system decoherence properties when the unavoidable interaction with the environment is included and temperature effects are considered.
\end{abstract}
\maketitle

\section{Introduction}
In solid state physics electrons or holes confined in 0-dimensional nanostructures, i.e. quantum dots (QDs), represent promising platforms for the realization of qubits, exploiting spin and/or charge, for quantum computing applications \cite{Loss-1998,DiVincenzo-2000,Veldhorst-2014,Morton-2011,Kawakami-2014,Ferraro-JPC2018}. However qubits are inevitably coupled to the degrees of freedom of the surrounding environment causing a loss of coherence that deeply affects the qubit operations \cite{Ferraro-QIP2018,Ferraro-ADV2018}. Depending on the nature of the host materials, a source of noise could be predominant with respect to the others. For example, materials belonging to group IV, such as Si and Ge, possess isotopes with zero nuclear spin that allow to reduce magnetic noise, while electrical noise remains an issue to be faced \cite{Simmons-2011,Xiao-2010}. 

The dynamics of electron spin in quantum dot is mainly affected by the interaction with two environments of different nature: the phonons in the lattice and the spins of atomic nuclei in the quantum dot. The spin-orbit interaction couples the spin and the orbital degrees of freedom that being coupled to the phonons, provide an indirect coupling between the electron spin and the phonons. They constitute a large dissipative bosonic reservoir causing decoherence and relaxation. The short-time correlations in the phonon bath induce a Markovian dynamics of the qubit that is the subject of our study. Moreover the electron spin and the nuclear spins in the host material interact via the Fermi contact hyperfine interaction, that creates entanglement between them. It turns out that long-time correlations in the nuclear spin system induce a non-Markovian dynamics of the electron spin. However the presence of silicon isotopes with zero nuclear spin reduce magnetic noise and non-Markovian dynamics becomes negligible.

Large progresses have also been done in studying semiconducting QDs in III-V compounds, such as GaAs \cite{Kikkawa-1998,Barthel-2010}, that assure greater advantages in fabrication processes; however Si qubits attracted recently a lot of attention also due to the immediate integrability with the existing CMOS technology of the microelectronic industry \cite{Rotta-2016, Rotta-2017}. 

When Si QDs based qubits are considered, the sixfold degeneracy of the conduction band minima, that is due to the two-fold degeneracy of the $\Delta$ valleys aligned along each one of the three main crystallographic directions, is an additional source of decoherence that may be overcome only if the typical qubit splitting energies are smaller with respect to the valley splittings \cite{DeMichielis-2012,Friesen-2010}. Otherwise, in order to have a complete picture, it becomes indispensable to include valley effects in the Hamiltonian model.

The hybrid qubit (HQ) is realized by the electrostatic confinement of three electron spins in a double quantum dot \cite{Shi-2012,Kim-2015}. We describe an HQ with an effective three-level model adopting a basis whose logical states are encoded in the $S=1/2$ and $S_z=-1/2$ subspace, where $S$ denotes the total angular momentum of the three electrons \cite{Ferraro-2014,Demichielis-2015}. Then, we model the environment with which the HQ unavoidably interacts, by a bath consisting of a series of harmonic oscillators with frequencies $\omega_j$. The effects of the bath temperature and of the bath spectral density on the qubit decoherence and relaxation times are studied.   

In Ref.\cite{Thorgrimsson-2017} the authors focus on ameliorating the dominant sources of decoherence in order to increase the coherence time in a Si/SiGe HQ. They measure $dE_Q/d\epsilon$, where $E_Q$ is the qubit energy and $\epsilon$ is the detuning between the two QDs and demonstrate that HQ can be made resilient to charge noise by tuning appropriately the qubit parameters. More recently in Ref.\cite{Abadillo-2018} atomic scale disorder at the quantum well interface is put into direct connection with the dephasing of the HQ.

The study of the relaxation and decoherence processes is of foundamental as well as practical interest for quantum computation applications. For this reason, the aim of the present paper is to study theoretically the phonon-induced relaxation and decoherence times and how these times are affected by the HQ parameters as well as by the bath structure.

The paper is organized as follows. In Section II we present the theoretical model describing the Si HQ including valley degeneracy, the bath and their interaction; moreover the relaxation and decoherence times are derived following Bloch-Redfield theory. Section III is devoted to the analysis of the relaxation and decoherence processes when the effects of the bath are included and the space of the qubit parameters is explored. Finally concluding remarks are reported in Section IV. 

\section{Theory}
This Section is devoted to the description, through an effective Hamiltonian model, of the HQ interacting with a bath of harmonic oscillators when temperature effects are included. The analytical expressions for the evaluation of relaxation and decoherence times are presented.
\subsection{Model of the silicon hybrid qubit in a thermal bath}
\begin{figure}[htbp]
\begin{center}
\includegraphics[width=0.3\textwidth]{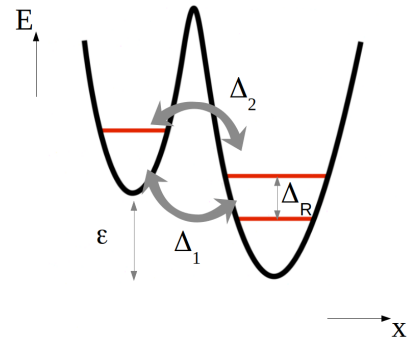}
\end{center}
\caption{A sketch of the HQ energy levels. The interdot tunnel couplings are $\Delta_1$ and $\Delta_2$, $\epsilon$ is the detuning between the two QDs and $\Delta_R$ corresponds to the low-energy splitting of the right dot.} \label{HQ}
\end{figure}

We describe effectively the HQ adopting a three dimensional basis. The first state of the basis corresponds to a configuration with two electrons in the left dot and one in the other and consequently has a singlet charge form. The remaining basis states, on the contrary correspond to the complementary configuration in which one electron is confined in the left dot and two electrons in right dot; they correspond to singlet and triplet charge configurations. The three-level matrix describing the qubit is
\begin{equation}\label{effmatrix2}
H_S =
\left( 
\begin{array}{ccc}
\frac{\epsilon}{2} & \Delta_1 & \Delta_2 \\
\Delta_1 & -\frac{\epsilon}{2} & 0\\
\Delta_2 & 0 & -\frac{\epsilon}{2}+\Delta_R
\end{array} 
\right),
\end{equation}
where $\epsilon$ is the detuning between the two QDs, $\Delta_1$ and $\Delta_2$ refer to the tunnel couplings between different charge states from one dot to the other and $\Delta_R$ corresponds to the low-energy splitting of the right dot, which reflect a valley excitation, an orbital excitation or a combination. The detuning $\epsilon$ can be changed by varying the applied electrostatic potential in the left QD. The tunnel couplings $\Delta_1$ and $\Delta_2$ can be electrostatically modulated by changing the tunneling barrier between the two QDs. $\Delta_R$ parameter can be effectively manipulated by varying the quantum confining energy profile of the right dot, exploiting external electric field coming from gate(s) close to the quantum dot. In particular, in the case of a Si/SiO$_2$ confining interface, the valley splitting depends on the electric field at that interface \cite{Yang-2013} so additional gate(s) (for example a back gate in a Silicon On Insulator structure) can be used to effectively modulate the electric field at the interface and the resulting $\Delta_R$ without affecting the confining energy potential. The estimation of such parameters is extractable from simulations adopting a tight-binding model as done in Ref.\cite{Abadillo-2018} for a strained Si quantum well sandwiched between strain-relaxed Si$_{0.7}$Ge$_{0.3}$. An illustrative sketch of the theoretical model describing HQ is reported in Fig.\ref{HQ}.

We model the surrounding environment by a series of $N$ harmonic oscillators with the Hamiltonian
\begin{equation}\label{HB} 
H_B=\sum_{j=1}^N\omega_jb_j^{\dagger}b_j,
\end{equation}
where $b_j(b_j^{\dagger})$ is the annihilation (creation) operator for the environment mode and $\omega_j$ is the frequency associated to each mode $j$.

The interaction Hamiltonian between HQ and the bath is written by \cite{Qin-2016}
\begin{equation}\label{HI} 
H_I=\sum_{j=1}^N\lambda_j(b_j^{\dagger}+b_j)\otimes\frac{1}{2}\hat{O}_S
\end{equation}
with $\lambda_j$ representing the coupling qubit-bath and the system operator $\hat{O}_S$ is equal to
\begin{equation}
\hat{O}_S=
\left( 
\begin{array}{ccc}
1 & 0 & 0 \\
0 & -1 & 0\\
0 & 0 & -1
\end{array} 
\right).
\end{equation}

The total Hamiltonian given by the sum of the three contributions, i.e. $H=H_S+H_B+H_I$, is transformed by adopting a unitary transformation $U=[\bold{m}_0, \bold{m}_1, \bold{m}_2]$. Each column of $U$ contains the eigenvectors $\bold{m}_k$ of $H_S$, in such a way that transforming $H_S$ through $U$, it results in a diagonal form. After making explicit calculations for $\tilde{H}=U^{\dagger}HU$, we finally obtain
\begin{equation}
\tilde{H}=
\left( 
\begin{array}{ccc}
E_0 & 0 & 0 \\
0 & E_1 & 0\\
0 & 0 & E_2
\end{array} 
\right)+\frac{\zeta }{2}\otimes
\left( 
\begin{array}{ccc}
\chi_{00} & \chi_{01} & \chi_{02} \\
\chi_{10} & \chi_{11} & \chi_{12}\\
\chi_{20} & \chi_{21} & \chi_{22}
\end{array} 
\right)+\tilde{H}_B,
\end{equation}
where $E_i$ with $i=0,1,2$ are the eigenvalues of $H_S$, $\zeta =\sum_{j=1}^N\lambda_j(b_j^{\dagger}+b_j)$, $\chi_{ij}=\chi_{ji}$ are the transformed matrix elements of $\hat{O}_S$ through $U$ and $\tilde{H}_B=H_B$.

\subsection{Relaxation and decoherence times}
We determine the explicit expressions for the relaxation and the decoherence times, firstly calculating the power spectrum $S_{\zeta}(\omega)$ for a bath in thermal equilibrium at temperature T. In the framework of the Linblad master equation describing the HQ, we trace over the environmental bath degrees of freedom, obtaining
\begin{equation}\label{S}
S_{\zeta}(\omega)=\frac{1}{2\pi}\int_{-\infty}^{+\infty}\langle f(t)f(0)\rangle_{\beta}\,e^{i\omega t}\,dt.
\end{equation}
The correlator $\langle f(t)f(0)\rangle_{\beta}$, where $\beta\equiv(k_BT)^{-1}$, is evaluated analytically giving as a result 
\begin{equation}\label{f}
\begin{split}
\langle f(t)f(0)\rangle_{\beta}&\equiv Tr_B(e^{-\beta H_B}f(t)f(0))\\
&=Tr_B(e^{-\beta H_B}e^{iH_Bt}\zeta e^{-iH_Bt}\zeta )\\
&=\sum_{j=1}^N\lambda_j^2\left[\cos\left(\omega_jt\right)\coth\left(\frac{\beta\hbar\omega_j}{2}\right)-i\sin\left(\omega_jt\right)\right],
\end{split}
\end{equation}
and the following relations have been exploited
\begin{align}
&\langle b_j^{\dagger}b_j\rangle_{\beta}=\frac{1}{e^{\beta\omega}-1}=\frac{1}{2}\coth\left(\frac{\beta\hbar\omega_j}{2}\right)-\frac{1}{2}\nonumber\\
&\langle b_jb_j^{\dagger}\rangle_{\beta}=\langle b_j^{\dagger}b_j+1\rangle_{\beta}=\frac{1}{2}\coth\left(\frac{\beta\hbar\omega_j}{2}\right)+\frac{1}{2}.
\end{align}
Inserting Eq. (\ref{f}) into Eq. (\ref{S}) we obtain
\begin{equation}\label{S2}
\begin{split}
S_{\zeta }(\omega)=&\frac{\sum_{j=1}^N\lambda_j^2}{2}\left\{\delta(\omega+\omega_j)\left[\coth\left(\frac{\beta\hbar\omega_j}{2}\right)-1\right]+\right.\\
&\left.+\delta(\omega-\omega_j)\left[\coth\left(\frac{\beta\hbar\omega_j}{2}\right)+1\right]\right\}.
\end{split}
\end{equation}
In the hypothesis that $N$ is large, the sum over $j$ can be approximated by a frequency integral $\sum_{j=1}^N\lambda_j^2\approx\int_{0}^{+\infty}J(\omega)\,d\omega$, where $J(\omega)$ is the spectral function of the oscillator bath, and Eq. (\ref{S2}) is rewritten in this way
\begin{align}\label{S3}
&S_{\zeta }(\omega)=\frac{1}{2}J(\omega)\left[\coth\left(\frac{\beta\hbar\omega}{2}\right)+1\right],\;\omega>0\\
&S_{\zeta }(\omega)=\frac{1}{2}J(\omega)\left[\coth\left(\frac{\beta\hbar\omega}{2}\right)-1\right],\;\omega<0.
\end{align}
The spectral density is supposed to be of the following general form: $J(\omega)=(\eta\omega^s/\omega_c^{s-1})e^{-\omega/\omega_c}$ with a high-energy cutoff $\omega_c$ and where $\eta$ is an effective dimensionless coupling that determines the overall strength of the electron-phonon coupling. The parameter $s$ distinguishes among $s=1$ Ohmic, $s>1$ super-Ohmic and $s<1$ sub-Ohmic baths.

Following the theory \cite{Borhani-2006,Kornich-2018}, in which the Bloch-Redfield master equation has been used to describe the dynamics of the qubit interacting with the phonon bath, the relaxation time $T_1$, the pure dephasing time $T_{\phi}$ and the decoherence time $T_2$ are directly linked to the power spectrum $S_{\zeta}(\omega)$ by the following relations
\begin{align}
&\frac{1}{T_1}=\frac{\pi}{2}\chi_{10}^2S_{\zeta }(E_Q)\label{eq1}\\
&\frac{1}{T_{\phi}}=\frac{\pi}{4}(\chi_{11}-\chi_{00})^2S_{\zeta }(0)\label{eq2}\\
&\frac{1}{T_2}=\frac{1}{2T_1}+\frac{1}{T_{\phi}},\label{eq3}
\end{align}
where $E_{Q}\equiv E_1-E_0$ is the qubit energy. The power spectrum $S_{\zeta}(0)$ is calculated at first order of the Taylor expansion giving as a result $S_{\zeta}(0)\approx\eta/(\hbar\beta)$ for the Ohmic regime and $S_{\zeta}(0)\approx\eta\omega/(\hbar\omega_c\beta)$ in the super-Ohmic regime. Then in the extreme sub-Ohmic case, $J(\omega)≈\eta\omega_{cutoff}$ (i.e. a nonzero constant at low frequency) that corresponds to $S_{\zeta }(0)\approx\eta\omega_{cutoff}/(\hbar\omega\beta)$.

\section{Results}
In this Section we report a detailed analysis on the relaxation and decoherence times when different experimental parameters related to the bath as well as to the HQ are varied.

\begin{figure}[htbp]
\begin{center}
\includegraphics[width=0.5\textwidth]{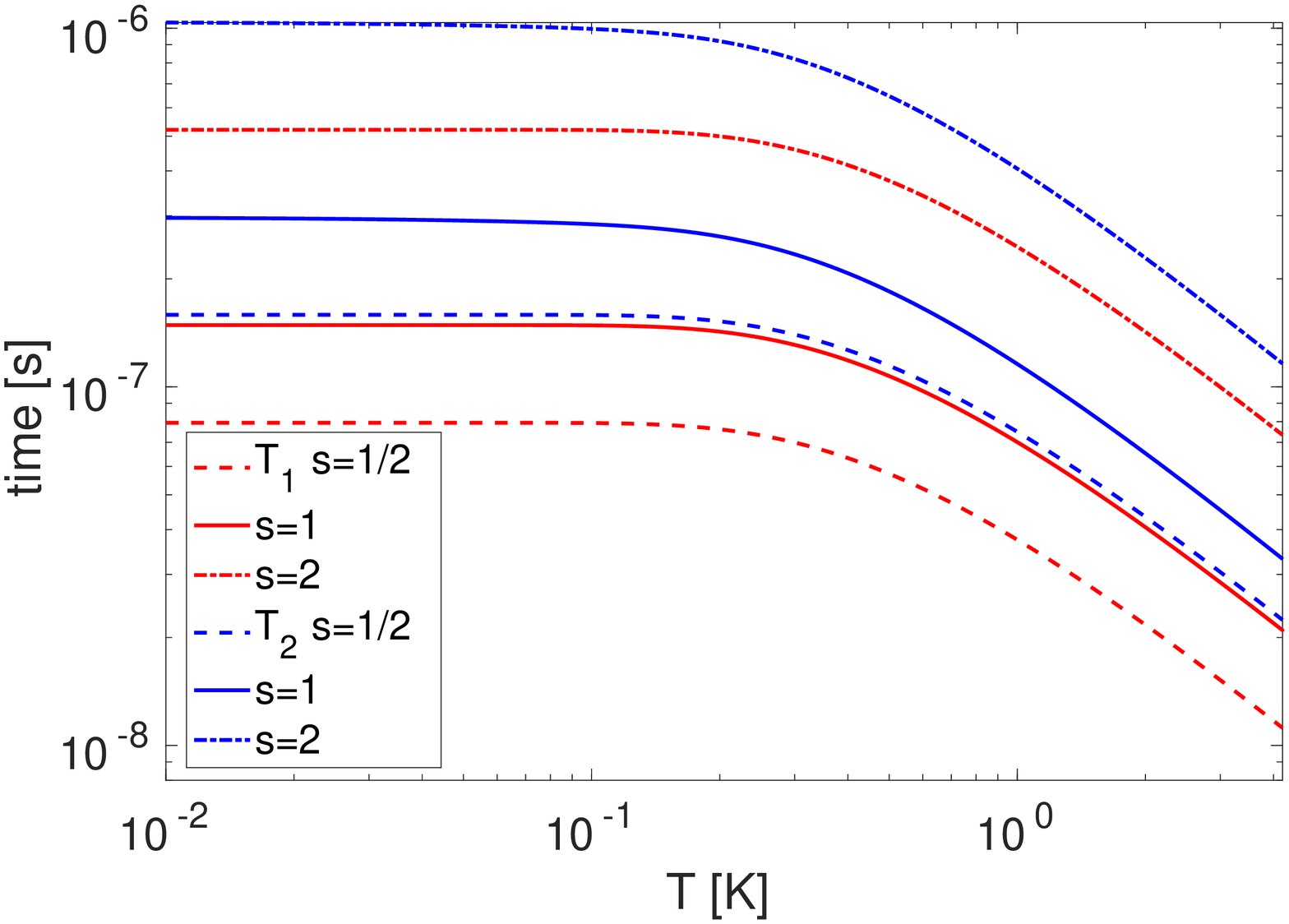}
\end{center}
\caption{$T_1$ (red lines) and $T_2$ (blue lines) as a function of the bath temperature T for three different regimes: $s=1$ (solid lines), $s=2$ (dot-dashed line) and $s=1/2$ (dashed line). The Si/SiGe HQ parameters are set to: $\epsilon=225\;\mu$eV, $\Delta_1=19.27$ $\mu$eV, $\Delta_2=12.20$ $\mu$eV, $\Delta_R=54.18$ $\mu$eV \cite{Abadillo-2018}. The bath parameters are set to: $\eta=0.5$, $\omega_c=10\Delta_1$, $\omega_{cutoff}/2\pi= 1$ Hz \cite{Cangemi-2018} .} \label{s_parameters}
\end{figure}

In Fig. \ref{s_parameters} the behaviour of the relaxation $T_1$ (red lines) and the decoherence $T_2$ (blue lines) times calculated through Eqs. (\ref{eq1})-(\ref{eq3}) is reported as a function of the bath temperature T for the three different regimes: $s=1$ Ohmic bath (solid lines), $s=2$ super-Ohmic bath (dot-dashed line) and $s=1/2$ sub-Ohmic bath (dashed line). The parameter $\eta$ is chosen in such a way to assure the Hamiltonian-dominated regime, that is $E_Q\gg\eta k_BT$, and at the same time to obtain relaxation and dephasing times compatible with experimental results recently obtained \cite{Thorgrimsson-2017} for the parameter values analyzed. The parameters of the HQ defined in Si/SiGe QDs as well as the bath parameters are taken from the literature \cite{Abadillo-2018,Cangemi-2018}. 

The relaxation time, as well as the decoherence, increase when the bath passes from a sub-Ohmic to a super-Ohmic regime and decrease when the bath temperature grows.

\begin{figure}[htbp]
\begin{center}
\includegraphics[width=0.52\textwidth]{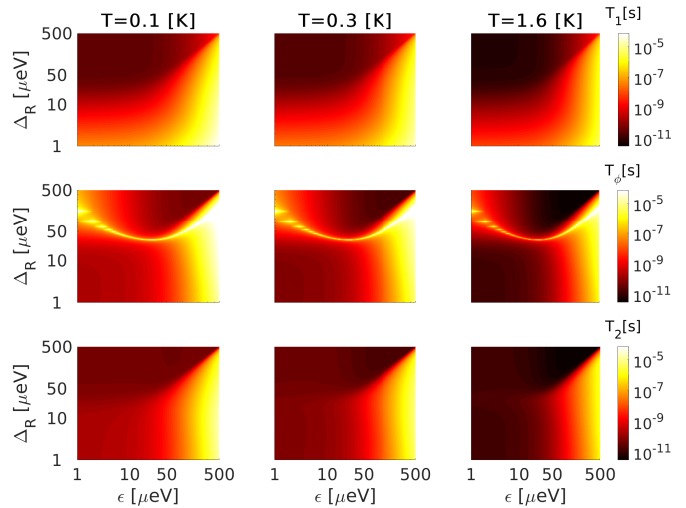}
\end{center}
\caption{$T_1$ (top), $T_{\phi}$ (middle) and $T_2$ (bottom) as a function of $\epsilon$ and $\Delta_R$ in correspondence to an Ohmic bath ($s=1$) at three different temperatures: T=0.1 K, 0.3 K and 1.6 K. The other qubit and bath parameters are the same as in Fig. \ref{s_parameters}.} \label{2D-T1T2}
\end{figure}

Guided by experimental results in which the coherence times are estimated in the range of hundreds of ns \cite{Thorgrimsson-2017,Abadillo-2018}, we choose to focus on the Ohmic regime. We analyse in Fig. \ref{2D-T1T2} the relaxation $T_1$, the pure dephasing $T_{\phi}$ and the decoherence $T_2$ times as a function of two significant qubit parameters that are the detuning $\epsilon$ that is tunable from external control voltages and the low-energy splitting of the right dot $\Delta_R$ that is linked to the qubit fabrication. We explore larger values of $\Delta_R$ with respect to the Si/SiGe case in order to include the valley splitting achievable in Si-MOS HQ \cite{Veldhorst-2014}. For the 2D plots we select three significant temperatures for experimentalists, that are T=0.1 K, 0.3 K and 1.6 K. 

As is seen in Fig. \ref{2D-T1T2}, the relaxation time increases when the detuning is large and, in the region where $\epsilon$ is smaller, the valley splitting $\Delta_R$ has to be keep small in order to assure larger times. Looking at the pure dephasing time, it increases in the large bias region and presents also high values in a narrow section for high $\Delta_R$ where ($\chi_{11} - \chi_{00}$) $\simeq$ 0 (see Eq. (\ref{eq2})). When the relaxation time is combined to the pure dephasing time, it then gives a smaller contribution to the total decoherence time in the large bias region than at narrow section. The overall result is that the decoherence time in the large bias region rises above its values at the narrow section albeit the latter remains a local section of maximum for the coherence time. All the characteristic times generally reduce as the temperature is increased.

\subsection{Relaxation time}
\begin{figure}[htbp]
	\begin{center}
		\includegraphics[width=0.3\textwidth]{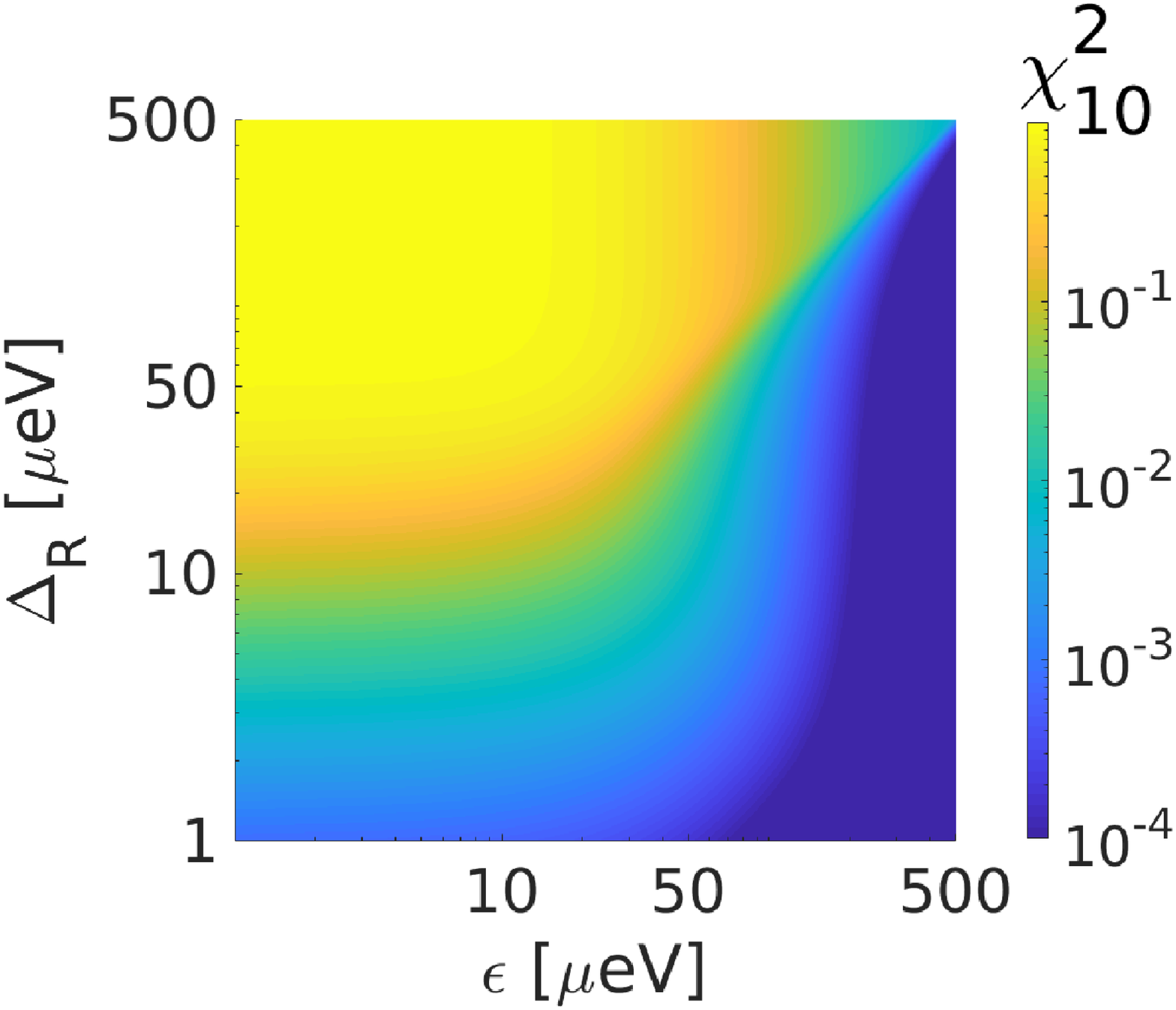}\\
                (a)
		\includegraphics[width=0.55\textwidth]{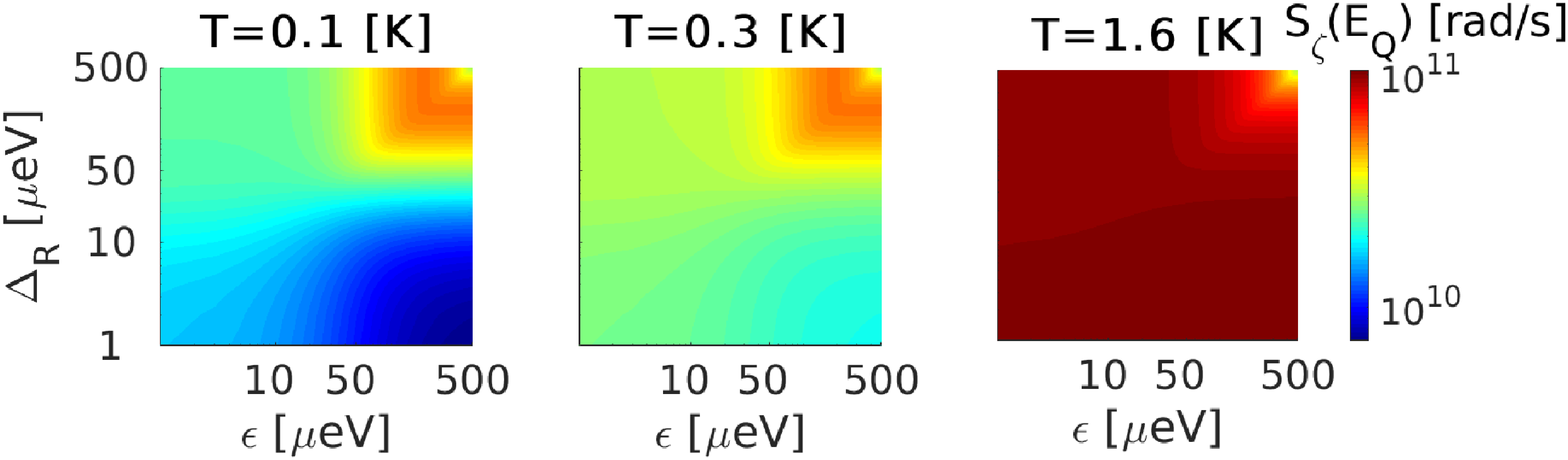}\\
                (b)
	\end{center}
	\caption{(a) $\chi_{10}^2$ as a function of $\epsilon$ and $\Delta_R$. (b) The power spectrum $S_{\zeta}(E_Q)$ in the same ($\epsilon$, $\Delta_R$) range of (a) for T=0.1 K, 0.3 and 1.6 K.} \label{Gamma1}
\end{figure}

We focus now our attention on $T_1$. In Fig. \ref{Gamma1} we report how the two ingredients composing the $T_1^{-1}$ behave against the detuning $\epsilon$ and $\Delta_R$: $\chi_{10}^2$ is plotted in Fig. \ref{Gamma1}(a) whereas the power spectrum of the bath $S_{\zeta}(E_Q)$ is presented in Fig. \ref{Gamma1}(b). Both the functions are calculated in the same range used in Fig. \ref{2D-T1T2} and $S_{\zeta}(E_Q)$ is presented for the three different temperatures considered, while $\chi_{10}^2$ depends uniquely on the qubit parameters.  

Note how $\chi_{10}^2$ can be heavily modulated in the range studied, suggesting that driving an HQ in the region of small $\epsilon$ and featuring high $\Delta_R$ has to be avoided to obtain high $T_1$. On the contrary, $S_{\zeta}(E_Q)$ does not present the same tunability, even if it depends on the $E_{Q}$ through $J(\omega)$ and $\coth(\hbar\omega/2kT)$ with $\omega$=$E_Q$/$\hbar$ (see Eq. (\ref{S3})). Obviously $S_{\zeta}(E_Q)$ augments as T is increased (see again Eq. (\ref{S3})), contributing to magnify the detrimental effects of $\chi_{10}^2$ on the relaxation times as highlighted in the corresponding plots of the first row in Fig. \ref{2D-T1T2}.   

\subsection{Pure dephasing and decoherence times}
When the working point is set by choosing a value for $\epsilon$, it is interesting to analyze how $T_{1}$, $T_{\phi}$ and $T_{2}$ are affected by the tunnel couplings, partially defined by the geometry of the HQ. 

Our study is focused on the effect of ($\Delta_2$, $\Delta_R$) variations on HQ characteristic times. $\Delta_1$, that can be less effectively modulated with respect to $\Delta_2$, produce small variations on $T_1$, $T_\phi$ and $T_2$ when compared to those caused by $\Delta_2$ and $\Delta_R$.

\begin{figure*}[htbp]
\begin{center}
(a)\includegraphics[width=0.45\textwidth]{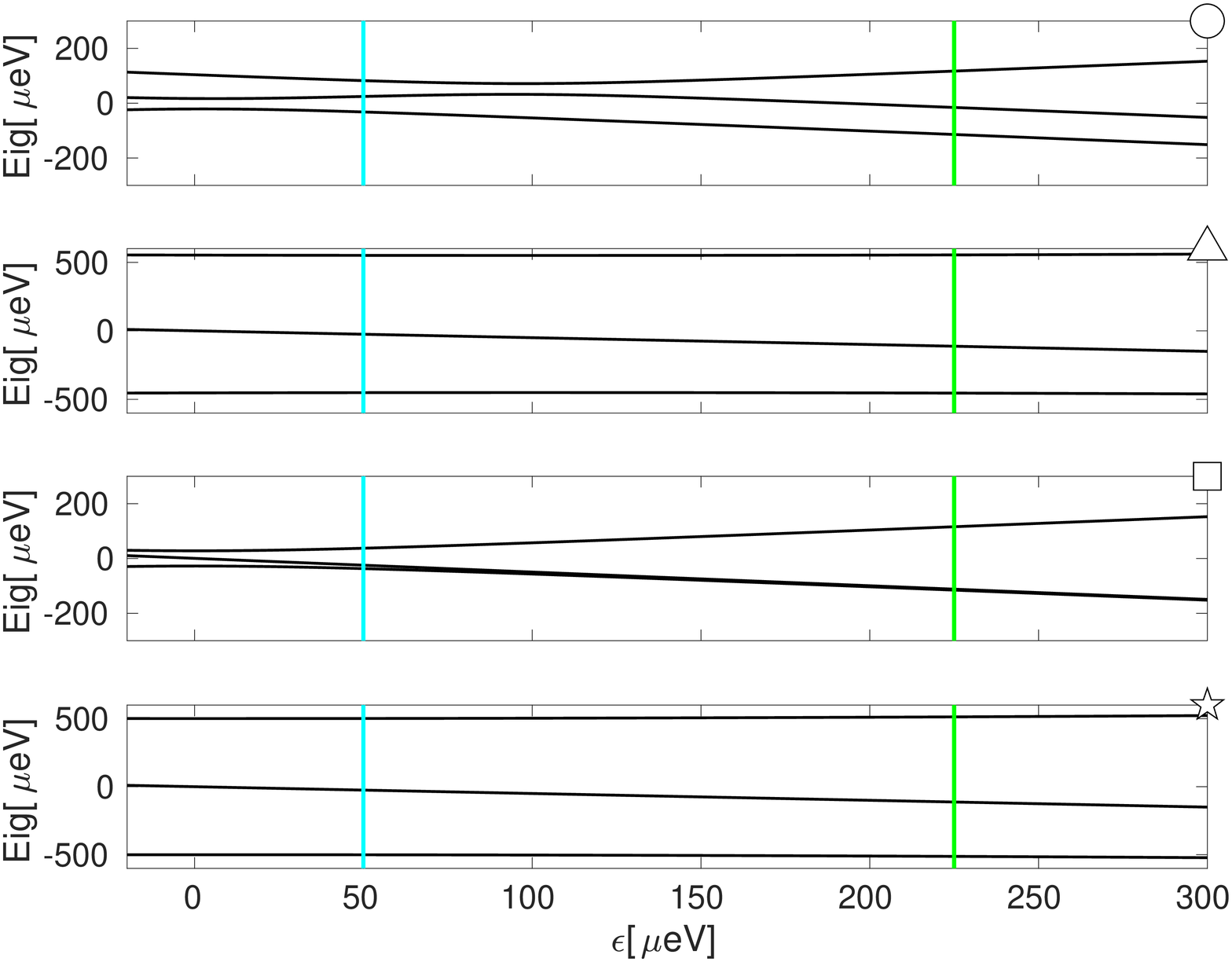}\
(b)\includegraphics[width=0.45\textwidth]{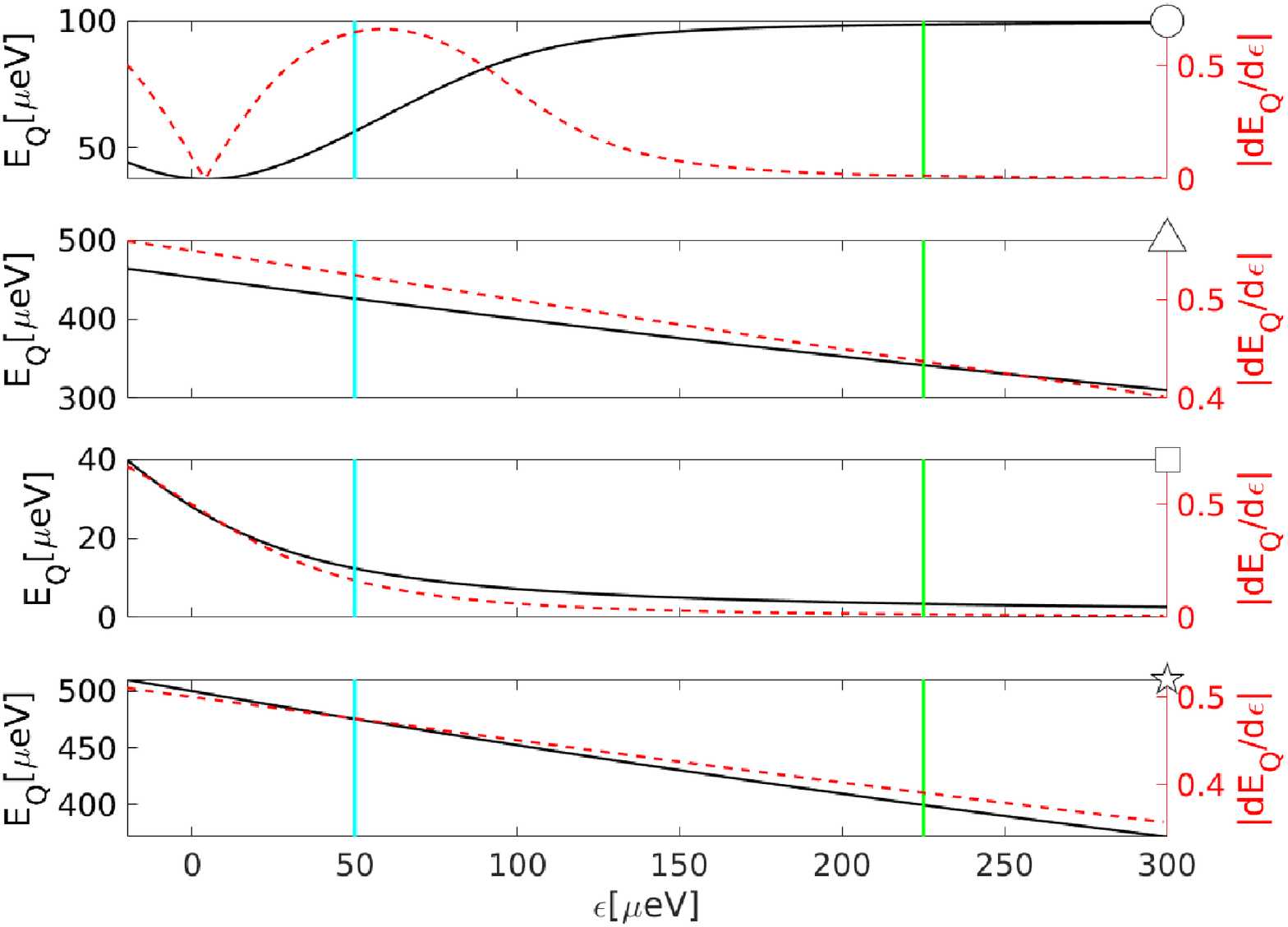}\\
(c)\includegraphics[width=0.45\textwidth]{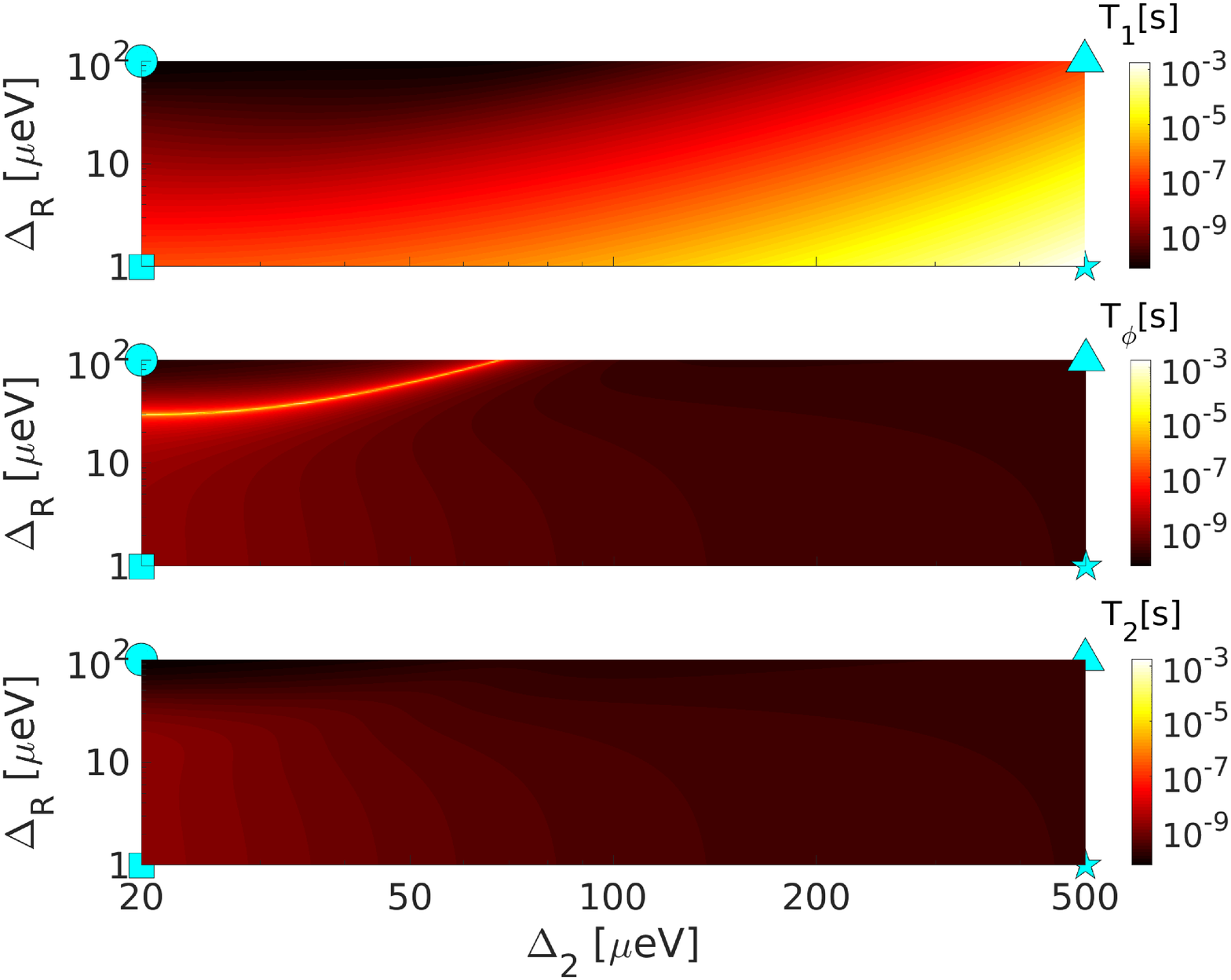}\
(d)\includegraphics[width=0.45\textwidth]{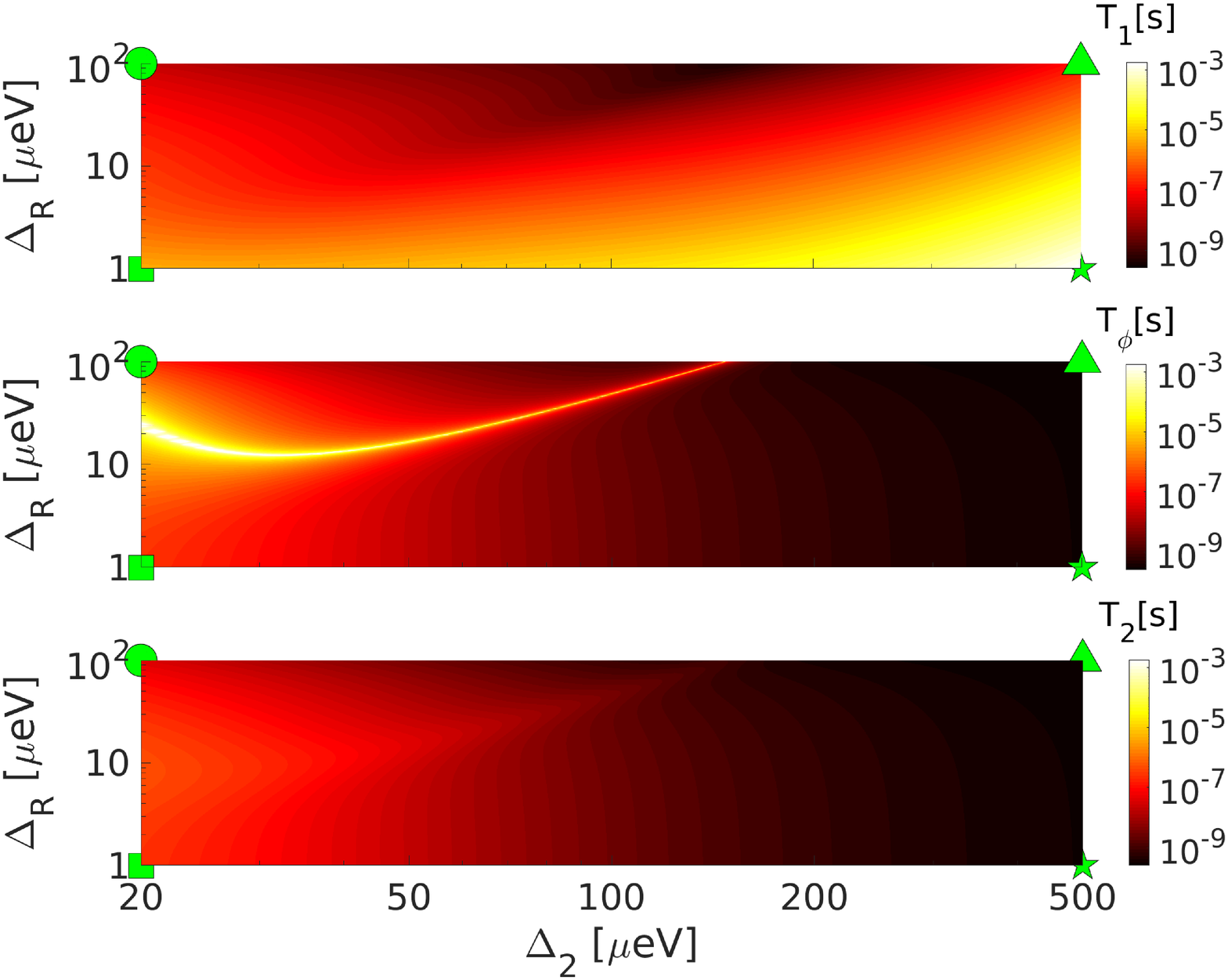}
\end{center}
\caption{(a) Eigenvalues of $H_S$ as a function of the detuning $\epsilon$ when $\Delta_2$ and $\Delta_R$ are set. The symbols at the corners of the subplots (circle, triangle, square and star) denote different qubit parameters (i.e. set of $(\Delta_2,\Delta_R)$ values) in correspondence to the range boundaries explored in (c) and (d). The colored vertical lines highlight the values of the detuning $\epsilon=50\mu$eV (cyan) and $\epsilon=225\mu$eV (green) chosen for the plots in (c) and (d) respectively. (b) Energy qubit $E_Q$ (solid black lines) and $dE_Q/d\epsilon$ (dashed red lines) as a function of $\epsilon$ for the same qubit parameter sets. (c) $T_1$ (top), $T_{\phi}$ (middle) and $T_2$ (bottom) as a function of $\Delta_2$ and $\Delta_R$ in correspondence to an Ohmic bath ($s=1$) at T=0.1 K and $\epsilon=50 \mu$eV. The other parameters are the same as in Fig. \ref{s_parameters}. (d) The same as (c) at $\epsilon=225 \mu$eV.} \label{2D-T1T2_eps}
\end{figure*}

To show how this analysis is strictly connected to the HQ eigenvalues trend, we plot in Fig. \ref{2D-T1T2_eps}(a) the eigenvalues of $H_S$ and in Fig. \ref{2D-T1T2_eps}(b) the qubit energy $E_{Q}$ (solid black line) and its derivative with respect to $\epsilon$, that is $d E_{Q}/d\epsilon$ (dashed red line), both as a function of the detuning. All these quantities are calculated in correspondence to four different sets of $\Delta_2$ and $\Delta_R$ at the range boundaries explored in Fig. \ref{2D-T1T2_eps}(c)-(d) and are marked in the plots with different symbols (circle, triangle, square and star). We also add two vertical lines highlighting the values of the detunings set to: $\epsilon=50 \mu$eV (cyan) and $225 \mu$eV (green), that are the values chosen in Fig. \ref{2D-T1T2_eps}(c)-(d) respectively. Fig. \ref{2D-T1T2_eps}(c)-(d) show 2D plots in which $T_1$, $T_{\phi}$ and $T_2$ are reported as a function of $\Delta_2$ and of the low-energy splitting of the right dot $\Delta_R$ at a fixed temperature T=0.1 K, that is of interest in experiments, for the two detunings chosen. We conclude that, independently of the value of the detuning, in correspondence to small value of the qubit energy we have smaller relaxation and coherence times. This correspond also to the condition in which $\Delta_2$ is small and quite closer to $\Delta_1$ (circle and square) for $T_1$ while for $T_{\phi}$ and $T_2$ to the condition in which $\Delta_2$ is larger than $\Delta_1$ (triangle and star).  

Focusing on $T_2$, we observe that it is marginally affected by variations of the qubit parameters in the low bias regime (Fig. \ref{2D-T1T2_eps}(c)), whereas the coherence time can be much more improved in the high bias regime (Fig. \ref{2D-T1T2_eps}(d)), especially for low values of $\Delta_2$ parameter. In fact, in this region (including green circle and square), 
$T_2$ is enhanced thanks to the rise of $T_{\phi}$ for $\epsilon$ points with low $|dE_Q/d\epsilon|$ values, as highlighted in Fig. \ref{2D-T1T2_eps}(b).

\begin{figure}[htbp]
\begin{center}
\includegraphics[width=0.5\textwidth]{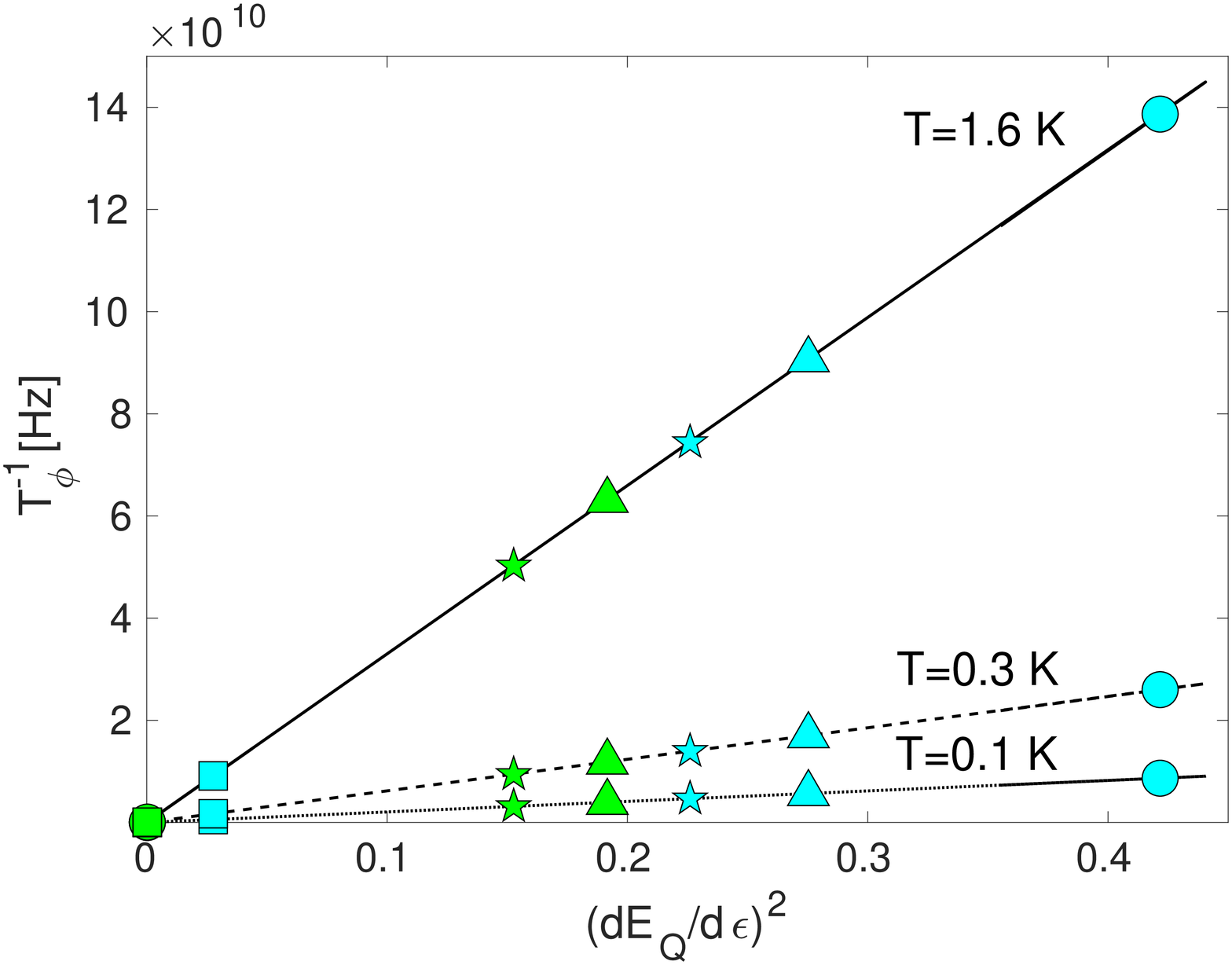}
\end{center}
\caption{Pure dephasing rate as a function of $(d E_{Q}/d\epsilon)^2$. The symbols corresponds to different qubit parameters highlighted in Fig. \ref{2D-T1T2_eps} (green symbols correspond to $\epsilon=50\mu$eV and cyan symbols to $\epsilon=225\mu$eV). The bath temperatures are: T=1.6 K (solid line), T=0.3 K (dashed line), T=0.1 K (dotted line).} \label{GammaPhi}
\end{figure}

To complete our analysis, we report in Fig. \ref{GammaPhi}, the pure dephasing rate $T_{\phi}^{-1}$ as a function of $(d E_{Q}/d\epsilon)^2$ for the different temperatures studied.

As it can be seen, the pure dephasing rate of the qubit shows a linear dependence on $(d E_{Q}/d \epsilon)^2$ with higher temperatures leading to higher slopes. Note that the configurations where $\epsilon$ assumes high values (green symbols) assuring low $dE_Q/d\epsilon$, produce lower dephasing rates than the cases with high $dE_Q/d\epsilon$, when $\epsilon$ is low (cyan symbols). This is due to the relation $(\chi_{11}-\chi_{00})\propto dE_Q/d\epsilon$ \cite{Qin-2016} that inserted in Eq. (\ref{eq2}) gives that $1/T_\phi\propto(dE_Q/d\epsilon)^2$.

\section{Conclusions}
The phonon-induced relaxation and decoherence processes are studied in the hybrid qubit in silicon quantum dots. We extract the relaxation, pure dephasing and decoherence times as a function of the bath spectral density and of the bath temperature using the Bloch-Redfield theory. For Si quantum dots the energy dispersion is strongly affected by the physics of the valleys so the contribution of the valley excitations has been effectively included in our analysis. It is found that the characteristics of both the spectral density of the bath and the energy spectrum of the qubit play an essential role. Contribution of phonons to relaxation and pure dephasing effects is bias dependent, leading to the conclusion that the coherence time can be higher in the large bias region than at the small bias, due to stronger relaxation at small bias. We also observed that the relaxation time is much more affected by the qubit energy spectrum than by the bath power spectrum. The pure dephasing rate exhibits a linear dependence on the square of the derivative of the qubit energy with respect to the detuning. This demonstrate a strong inverse proportionality, small values of the derivative of the qubit energy correspond to larger dephasing times, as confirmed by the experiments. Moreover the higher the temperature the higher the slope, meaning that at higher temperature there is a large variability of the pure dephasing time with respect to energy qubit derivative.

\acknowledgements
This work is partially supported by the European project H2020-ICT-2015 MOS-QUITO, Grant Agreement No. 688539.

\bibliography{Ref}

\end{document}